\documentclass[10pt,twocolumn,english,aps,prl,
amsmath,amssymb,colorlinks=true,linkcolor=blue,
citecolor=blue,urlcolor=blue]{revtex4-1}
\usepackage[T1]{fontenc}
\usepackage[latin9]{inputenc}
\usepackage{array}
\usepackage{float}
\usepackage{textcomp}
\usepackage{multirow}
\usepackage{amsmath}
\usepackage{graphicx}

\usepackage{xcolor}
\usepackage[colorlinks=true,linkcolor=blue,citecolor=blue,urlcolor=blue]{hyperref}

\makeatletter

\usepackage{graphicx,epsfig}
\usepackage{longtable}
\usepackage{xcolor}

\usepackage[caption=false]{subfig}

\makeatother

\usepackage{babel}

\begin{document}

\title{Expanding scroll rings in a model for the photosensitive Belousov-Zhabotinsky reaction}

\author{Arash Azhand\textonesuperior{}, 
 Rico Buchholz\texttwosuperior{},
 Jan F. Totz\textonesuperior{}, Harald Engel\textonesuperior{}}
\email{azhand@itp.tu-berlin.de}
\address{\textonesuperior{}Institut für Theoretische Physik, Technische Universität
Berlin, Hardenbergstrasse 36, D-10623 Berlin, Germany\\
\texttwosuperior{}Theoretische Physik V, Universität Bayreuth, Univeristätsstrasse
30, D-95440 Bayreuth, Germany}

\begin{abstract}
While free scroll rings are non-stationary objects that either grow or contract with time, 
spatial confinement can have a large impact on their evolution reaching from significant 
lifetime extension [J. F. Totz , H. Engel, and O. Steinbock, \href{http://iopscience.iop.org/article/10.1088/1367-2630/17/9/093043/meta;jsessionid=EB7AEAE850B3919E27D6F7325FEE1081.c4.iopscience.cld.iop.org}
{\textit{New J. Phys.} \textbf{17}, 093043 (2015)}]
up to formation of stable stationary and breathing pacemakers [A. Azhand, J. F. Totz, and H. Engel, 
\href{http://iopscience.iop.org/0295-5075/108/1/10004}{\textit{Europhys. Lett.} \textbf{108}, 10004 (2014)}]. Here, we explore the parameter range in which the 
interaction between an axis-symmetric scroll ring and a confining planar no-flux boundary 
can be studied experimentally in transparent gel layers supporting chemical wave propagation 
in the photosensitive variant of the Belousov-Zhabotinsky medium. Based on full three-dimensional 
simulations of the underlying modified complete Oregonator model for experimentally realistic 
parameters, we determine the conditions for successful initiation of scroll rings in a phase diagram  
spanned by the layer thickness and the applied light intensity. Furthermore, we discuss whether 
the illumination-induced excitability gradient due to Lambert-Beer's law as well as a possible 
inclination of the filament plane with respect to the no-flux boundary can destabilize the scroll ring.
\end{abstract}

\pacs{Valid PACS appear here}

\maketitle


\section{Introduction}

Undamped propagation of nonlinear excitation waves is an important example for self-regulated pattern formation in spatially extended biological, chemical, and physical systems. In particular rotating spiral waves have been observed in the Belousov-Zhabotinsky (BZ) medium \cite{Winfree:Science:72}, on the chicken retina 
\cite{Goroleva:JNeurobiol:1983}, on platinum surfaces \cite{Jakubith:PhysRevLett:90}, in slime molds 
\cite{Siegert:PhysicaD:91}, in the heart \cite{Davidenko:Nature:92}, in a liquid crystal system 
\cite{Frisch:PRL:1994}, in collonies of giant honey bees \cite{Kastberger:PLOS-One:2008}, and recently as traveling precipitation waves \cite{Tinsley:JPhysChemA:2013}. Scroll waves (SWs) are the natural extension of spiral waves in the third dimension. 

Here, we are interested in SWs with closed filaments, so called scroll rings (SRs). The first experimental observation of SRs was reported by Winfree \cite{Winfree:Science:73} in the framework of the BZ reaction. Later SRs were also observed in fibrilating cardiac tissue \cite{MedvinskyPanfilovPertsov1984}, for example.

The dynamics of a free SR in an unbounded medium (i.e., far from any boundaries) is defined by the filament tension $\alpha$. 
The filament contracts or expands with time depending on whether $\alpha$ is positive or negative. In the first case, 
the SR shrinks and usually collapses after a finite lifetime. In the second case, the ring grows and undergoes the 
negative line tension instability (NLTI) \cite{PanfilovRudenko1987, Bansagi:PhysRevE:07}, before eventually a state 
called vortex or Winfree turbulence developes \cite{Biktashev:IntJBifurcatChaos:1998, Alonso:Science:03a}. Thus, 
free SRs represent in-stationary objects for any value of $\alpha$. Several experimental \cite{Winfree:SciAm:74,
Welsh:Nature:1983, WinfreeJahnke1989, Bansagi:PhysRevLett:2006} and theoretical \cite{Yakushevich:1984,Keener:PhysicaD:88, 
Biktashev:PhilTransRSocA:94, Dierckx:PRE:2013} studies focused on the evolution of SWs in unbounded media.

The described above behavior of free SRs can fundamentally change under spatial confinement. Based on numerical simulations with the FitzHugh-Nagumo (FHN) model, Nandapurkar and Winfree argued already in 1989 that the interaction with a confining no-flux 
boundary (NFB) can stabilize an expanding SR \cite{Nandapurkar:PhysicaD:89}. Instead of undergoing the NLTI, the ring 
transforms in a three-dimensional autonomous pacemaker (3d APM). Recently, this new type of self-organized pacemaker 
has been observed experimentally in a photosensitive Belousov-Zhabotinsky (PBZ) medium with negative $\alpha$ \cite{azhand:epl:2014}. 
For the opposite case, $\alpha > 0$, in experiments with the ferroin-catalyzed BZ system a substantial increase of 
the SRs lifetime including very long-living transients have been reported for SRs in interaction distance to a planar NFB \cite{Totz:2015}.
Other effects of spatial confinement on scroll wave dynamics have been discussed by Dierckx et al. \cite{Dierckx:PRL:2012}, Biktasheva et al.
\cite{Biktasheva:PRL:2015}, and by Ke and colleagues \cite{Ke:Chaos:2015}.

\begin{figure*}
\begin{centering}
\includegraphics[width=1\linewidth]{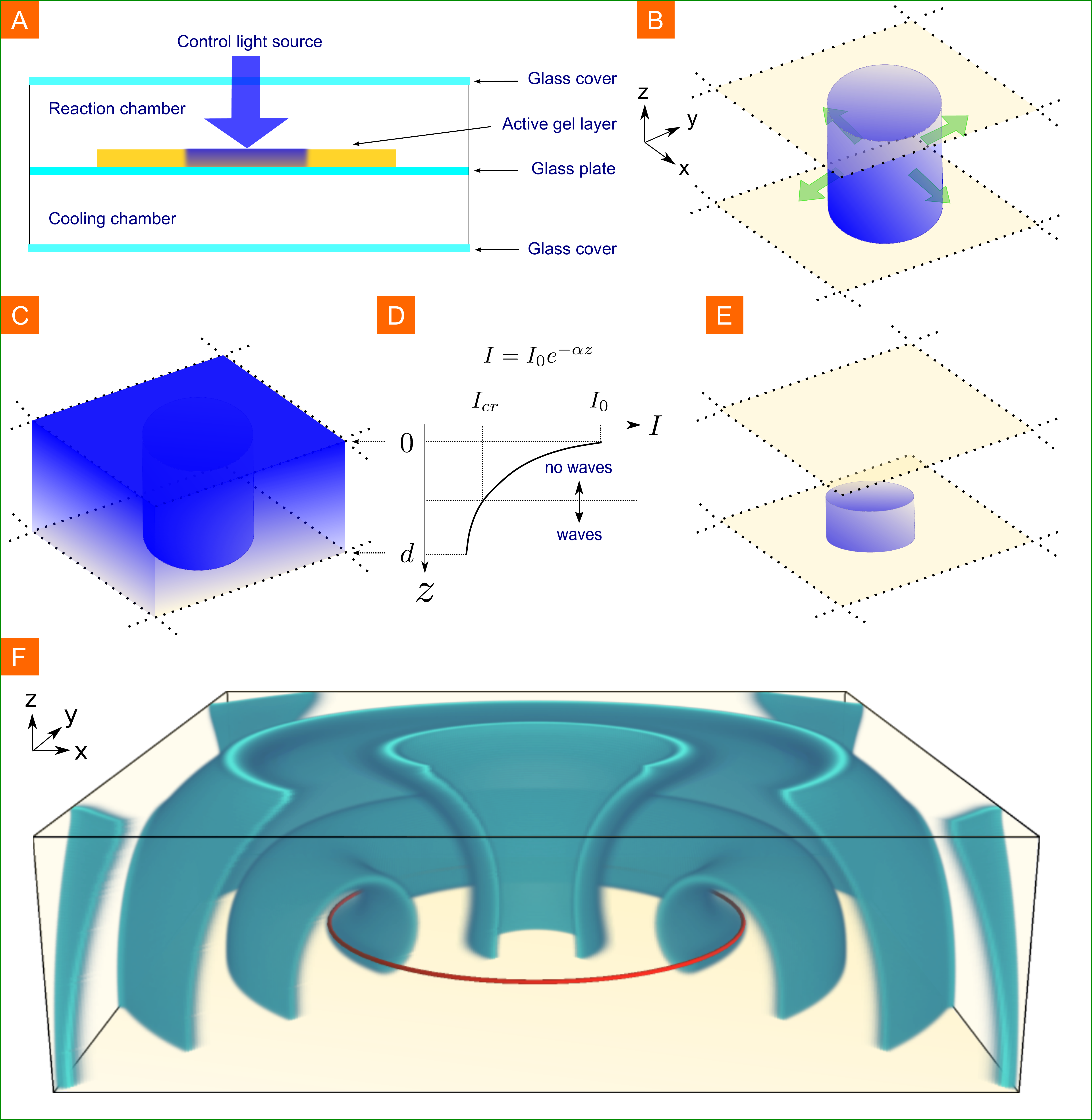}
\par\end{centering}

\caption{(Color online) Optical initiation of a SR in the PBZ reaction (schematic). (A) Cross-section of a two-chamber open gel reactor with the catalyst-loaded gel layer illuminated from above.
(B) Outward propagating cylindrical wave front (blue), green arrows indicate direction of propagation.
(C) Illumination of the layer from above with actinic light of intensity  $I_0$ turned on. 
(D) Attenuation of light intensity $I$ in $z$-direction according to Lambert-Beer's law. The wave front is gradually inhibited in the upper part of the layer where $I(z) > I_\text{cr}$. Here, $I_\text{cr}$ denotes the threshold beyond which the medium ceases to support wave propagation and $d$ is the layer thickness.
(E) After the light is turned off, the upper part of the gel layer recovers. Now, the open rim of the truncated cylindrical wave curls up forming a SR. 
(F) Fully developed axis-symmetric SR with filament plane in interaction distance to the lower NFB at $z = d$. The iso-concentration surface of the activator is shown in blue, the red line marks the  ring-shaped filament. The influence of the opposite NFB at $z = 0$ and the lateral boundaries is vanishingly small because the wave fronts in between filament and boundaries shield the interaction completely.  \label{fig:SimSpaceConditions}}
\end{figure*}

This paper is aimed at critically examine by realistic 3d numerical simulations to what extent experiments performed in a thin transparent 
gel layer loaded with a photosensitive BZ catalyst are suited to reveal the interaction of SWs with NFBs. The PBZ system is particularly 
promising as it allows for reproducible initiation of SRs with a filament plane in a certain desired distance to the NFB. The idea is to extinguish 
a cylindrical chemical wave partially by temporary intense illumination with actinic light. This idea is illustrated in Fig. \ref{fig:SimSpaceConditions}, 
showing schematically the open gel reactor (panel A) and the initiation of a SR by application of the actinic light (panels B to F). After the light 
is switched off, the truncated wave will curl up in the recovering medium and form the SR. A similar technique has been successfully applied previously, 
see \cite{Linde:PhysicaD:1991, Amemiya:PRL77:1996, Amemiya:Chaos:98}, for example. 
Because the interaction between spiral waves and NFBs is known to be short-range and screened \cite{Aranson:PhysRevE:1993, Aranson:PhysRevE:1994, 
Brandtstaedter:CPL:2000, HenryHakim:PhysRevE:2002, Biktasheva:PhysRevE:2003, Biktasheva:PhysRevE:2009, Kupitz:JPhysChemA:2013}, already for distances 
at the order of the wave length a two-dimensional spiral wave does not feel a nearby NFB. 
Therefore, gel layers utilized in open gel reactors for the PBZ reaction are sufficiently thick to position a SR in interaction distance to a NFB while simultaneously the 
interaction with the opposite and lateral boundaries is negligible (compare schematic sketch of the BZ setup in Fig. \ref{fig:SimSpaceConditions}).

The structure of the paper is as follows: Our full 3d numerical simulations are based on the modified complete Oregonator (MCO) model for the PBZ reaction developed by 
Krug et al. \cite{Krug:JPhysChem:90} which  will be recapitulated in section \ref{sec:Model}. In section \ref{sec:Scroll-ring-unbounded}, we briefly 
describe the evolution of a free SR for experimentally realistic recipe concentrations corresponding to negative line tension. Section \ref{sec:Scroll-ring-bounded} 
contains the results obtained for a SR that interacts with a confining planar NFB. The parameter range for successful optical 
initiation of SRs is calculated in section \ref{sec:Height_Vs_Lightintensity}. In this section we also discuss, whether the SR can be destabilized by the parameter gradient due to attenuation of light as described by Lambert-Beer's law. Finally, we address a possible destabilization of the SR caused by small inclination of the filament plane 
with respect to the NFB due to non-uniform illumination during SR initiation.

\section{Model\label{sec:Model}}

To study the evolution of SRs in thin layers of PBZ media numerically, we perform full 3d numerical simulations with using the modified complete Oregonator (MCO) model. This modification of the  Oregonator model proposed by Field, Körös, and Noyes was developed by Krug et al. to account for the inhibitory effect of light and/or oxygen on the reaction \cite{Krug:JPhysChem:90,Kadar:JPhysChemA:97a}. The MCO equations read: 

\begin{align}
\frac{\partial u}{\partial t} & =\frac{1}{\epsilon_{u}}\left[u-u^{2}+w(q-u)\right]+\nabla ^2 u,\nonumber \\
\frac{\partial v}{\partial t} & =u-v,\label{eq:MCO}\\
\frac{\partial w}{\partial t} & =\frac{1}{\epsilon_{w}}\left[\phi+fv-w(q+u)\right]+\delta \nabla ^2 w, \nonumber 
\end{align}

Here, the variables $u$, $v$ and $w$ are proportional to concentrations $\left[ HBr\ensuremath{O_{2}}\right]$ of bromous acid (activator), 
$\left[ Ru\ensuremath{\left(bpy\right)_{3}^{3+}} \right]$ of the oxidized catalyst, and $\left[ B\ensuremath{r^{-}}\right]$ of bromide (inhibitor), respectively. 
$\nabla ^2$ denotes the 3d Laplacian operator. The ratio of the diffusion coefficients is $\delta \, =\, D_{w}/D_{u}\, =\, 1.12$. 
There is no diffusive term in the equation for $v$ since in experiments the catalyst $Ru\left(bpy\right)_{3}^{3+}$ is immobilized in the gel layer. 

For the definition of the parameters in eqs. (1) as the recipe-dependent intrinsic kinetic time scales $\epsilon_{u}$ and $\epsilon_{w}$ with $\epsilon_{u}\gg\epsilon_{w}$, the ratio of rate constants $q$, 
the stochiometric parameter $f$ compare \cite{Krug:JPhysChem:90}. The photochemically induced bromide flow $\phi$ is assumed to be proportional to the applied light intensity. This parameter plays a crucial role in the experiments with the PBZ reaction because it specifies whether the nonlinear chemical kinetics is bistable, excitable or oscillatory. Varying the applied light intensity allows to control the local excitation threshold, the rotation regime of spiral waves forming a SR, and its filament tension, for example.

For our numerical investigations we have chosen the following parameter set: $\epsilon _u = 0.07$, $\epsilon _w = \epsilon / 90$, $f = 1.16$, $q = 0.002$, and $\phi = 0.014$. 
A two-dimensional spiral wave at this parameter set is performing rigid rotation with a rotation period $T=6.9$ and a wavelength of $\lambda=19.6$. Space and time are non-dimensionalized. 

Simulations are conducted with a forward Euler scheme for time integration and a nineteen point star discretization for the Laplacian. For space and time discretization we use 
$dx=dy=dz=0.3$ and $dt=0.001$, respectively. Spatial dimensions are $4\times4\times2\lambda^{3}$ for unbounded SRs and $4\times4\times1\lambda^{3}$ for confined SRs. NFB conditions 
($\nabla c = 0$ with $\nabla = \left( \partial_{x},\, \partial_{y},\, \partial_{z} \right) ^{T}$ and $c\in \left\{ u,\, v,\, w\right\}$) are chosen for all sides of both 
the bounded and the unbounded media.

Initiation of SRs in numerical simulations is accomplished as in the chemical experiment. First,  outward propagating cylindrical waves of different radii are generated as shown in (Fig. \ref{fig:SimSpaceConditions} B). Then, at a certain simulation time $t=0$, in the upper part of the layer the photochemically induced bromide flow $\phi$ is set to some value larger than the critical one for wave extinction. After resetting $\phi$ to the background value the SR developes as described in (Fig. \ref{fig:SimSpaceConditions} E) and evolves into a perfectly planar SR (Fig. 
\ref{fig:SimSpaceConditions} F). 

In our simulations the filament is defined as the intersection of two iso-concentration surfaces, given by $u_{i}=0.3$ and $v_{i}=0.1$, respectively. Typically, this definition yields an oscillation of 
both the filament position and the filament radius in time. This is due to the rigid rotation of the corresponding spiral waves.

\section{Scroll ring dynamics in uniform unbounded media\label{sec:Scroll-ring-unbounded}}

\begin{figure*}
\begin{centering}
\includegraphics[width=1\linewidth]{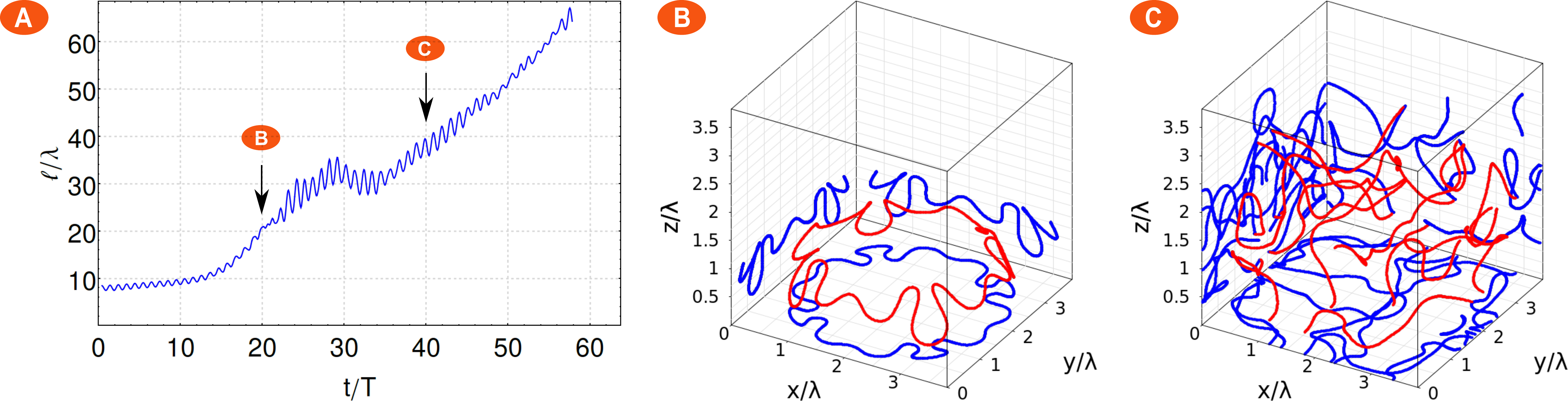}
\par\end{centering}

\caption{(Color online) (A) Growth of filament length of a free SR with negative filament tension. Black arrows indicate time moments of 20 and 50 periods of rotation after initiation, respectively, for which snapshots of filament evolution are presented in panels (B) and (C). Parameter values 
are $\epsilon_{u}=0.07$, $\epsilon_{w}=\epsilon_{u}/90$, $f=1.16$, $q=0.002$, and $\phi=0.014$.\label{fig:SRFreeMultView}}
\end{figure*}

Far from any boundary, the dynamics of axis-symmetric SRs is governed by two ordinary differential equations (ODEs) for the filament radius $R$ and the position of the filament plane $z$ 
\cite{Keener:PhysicaD:88,Biktashev:PhilTransRSocA:94}:

\begin{align}
\frac{dR}{dt} & =-\frac{\alpha}{R},\label{eq:filament_ode_free_one}\\
\frac{dz}{dt} & =\frac{\beta}{R}.\label{eq:filament_ode_free_two}
\end{align}

The constant coefficients $\alpha$ and $\beta$ are specified by the properties of the chosen excitable medium, i.e., by the parameters of the MCO model (1). Solutions to Eqs. (\ref{eq:filament_ode_free_one}) and (\ref{eq:filament_ode_free_two}), supplemented by the initial conditions $R(t=0)=R_0 >0$ and $z(t=0)=z_0$, are given by

\begin{align}
R(t) & =\sqrt{R_{0}^{2}-2\alpha t},\label{eq:fitfilament_R}\\
z(t) & =z_{0}-\frac{\beta}{\alpha}R(t).\label{eq:fitfilament_Z}
\end{align}

The filament tension $\alpha$ determines whether the SR shrinks ($\alpha>0$) or expands ($\alpha<0$) in the course of time. While a SR with positive filament tension will collapse within finite time,
in case of negative filament tension there will be an unbounded increase of the filament length accompanied by filament break-up.
Superimposed on radial expansion or contraction the SR drifts along its symmetry axis where the drift direction depends on $\beta$.

\begin{figure}[!t]
\begin{centering}
\includegraphics[width=1\columnwidth]{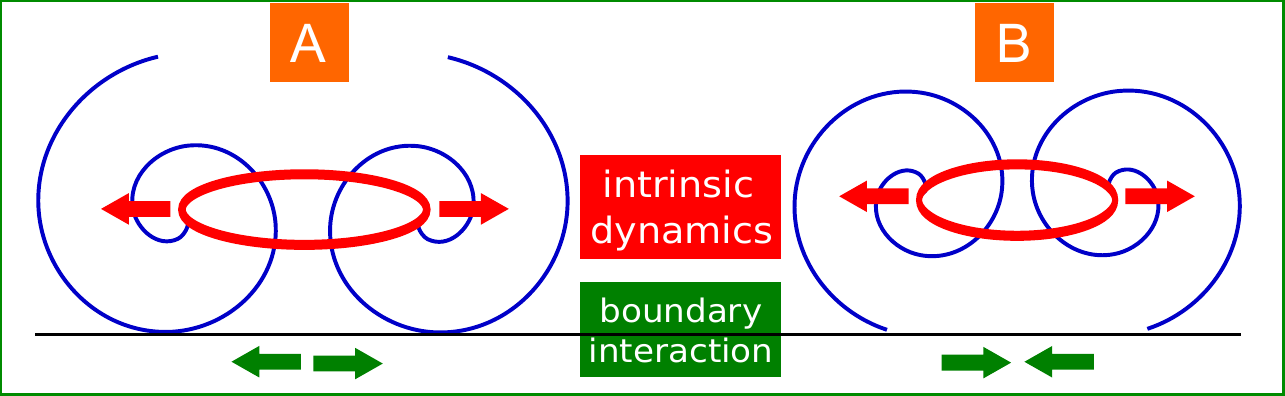}
\par\end{centering}

\caption{(Color online) Cooperative (A) and antagonistic (B) setting for the interaction of a SR  with a planar NFB (schematic). Arrows indicate the direction of intrinsic (red) and boundary-induced (green) radial dynamics. Spiral waves forming the ring rotate clockwise in (A) and counterclockwise in (B).\label{fig:SketchCoopAntag}}
\end{figure}

We calculated filament tension $\alpha$ and drift coefficient $\beta$ for the free SR by fitting Eqs. \eqref{eq:fitfilament_R} and \eqref{eq:fitfilament_Z} to numerical data. For the chosen parameter set (compare II) we found $\alpha = -0.03 \, \frac{\lambda ^{2}}{T}$ and $\beta = 0.005 \, \frac{\lambda ^{2}}{T}$ in units of the wavelength and rotation period of the spiral waves forming the ring.

The evolution of a sample free SR is shown in figure \ref{fig:SRFreeMultView}. Panel A displays the  the overall filament length vs. time. The circular filament grows and undergoes the NLTI (panel B). As a consequence the filament breaks up into single pieces that because of the NLTI grow further on its own. The emerging final state is known as vortex or Winfree turbulence (panel C) \cite{Biktashev:IntJBifurcatChaos:1998, Alonso:Science:03a}.

\section{Scroll ring interacting with a planar no-flux boundary\label{sec:Scroll-ring-bounded}}

\begin{figure*}
\begin{centering}
\includegraphics[width=1\linewidth]{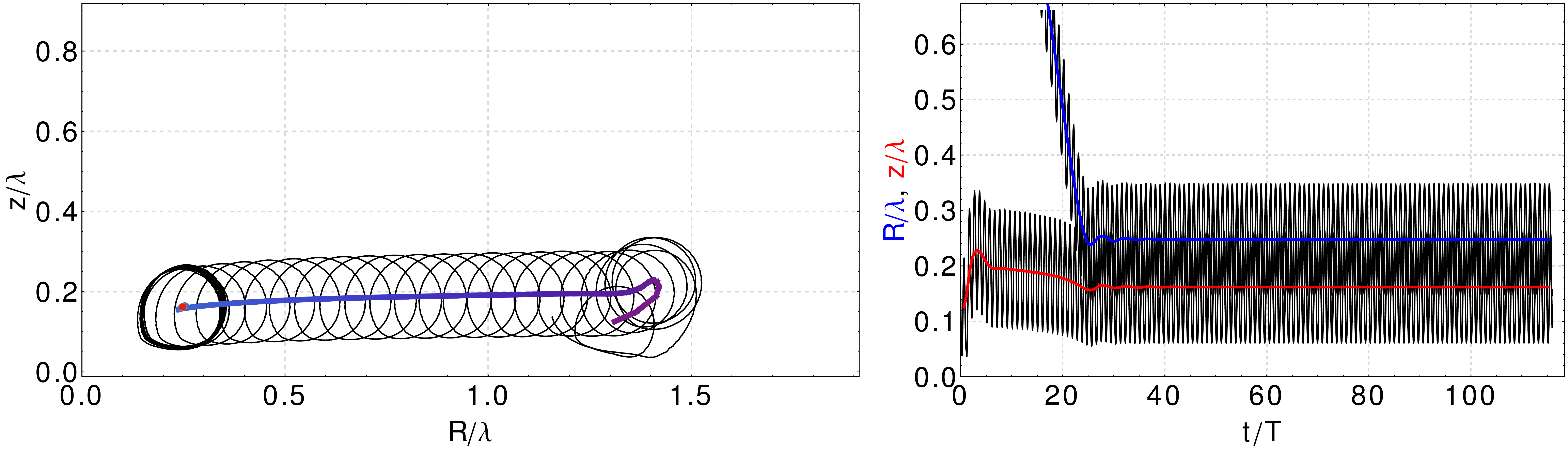}
\par\end{centering}

\caption{(Color online) 3d-numerical simulation of the evolution of a planar SR with symmetry axis perpendicular to the confining NFB. Parameters of the MCO model as in Fig. 2. Initial radius and distance between filament plane and NFB $R_0 = 1.4\, \lambda$ and $z_0 = 0.15\, \lambda$, respectively. \label{fig:BoundZandRoverTime-Par2}}
\end{figure*}


The interaction of an expanding SR with a NFB can either increase or decrease the radial growth of the filament ring.


 Panel A of Fig. \ref{fig:SketchCoopAntag} shows a setting where both intrinsic radial growth and boundary interaction act cooperativly leading to an amplified radial expansion of the ring. Contrary to cooperative setting, intrinsic dynamics and boundary effects can counteract each other. In this antagonistic setting illustrated in Fig. \ref{fig:SketchCoopAntag}, Panel B, an expanding free SR initiated close to a NFB can even contract actually. This possibility is exemplified in Fig. 
 \ref{fig:BoundZandRoverTime-Par2}. The SR shown would expand in an unbounded medium. Now, in interacting distance to the NFB, after a transient of roughly 20 rotation times $T$, a boundary-stabilized stationary SR forms with $R_{\infty} \sim 0.25\, \lambda$ at a distance $z_{\infty} \sim 0.16 \, \lambda$ from the lower boundary. This SR remains stable over at least $400$ periods despite of $\alpha < 0$, because the NLTI is suppressed by the interaction with the NFB (for the sake of clarity only the first $150$ periods are presented in Fig. \ref{fig:BoundZandRoverTime-Par2}). Here $R$ and $z$ are values averaged over rapid oscillations in the position of the filament due to rigid rotation of spiral waves forming the ring.

\begin{figure*}
\begin{centering}
\includegraphics[width=1\linewidth]{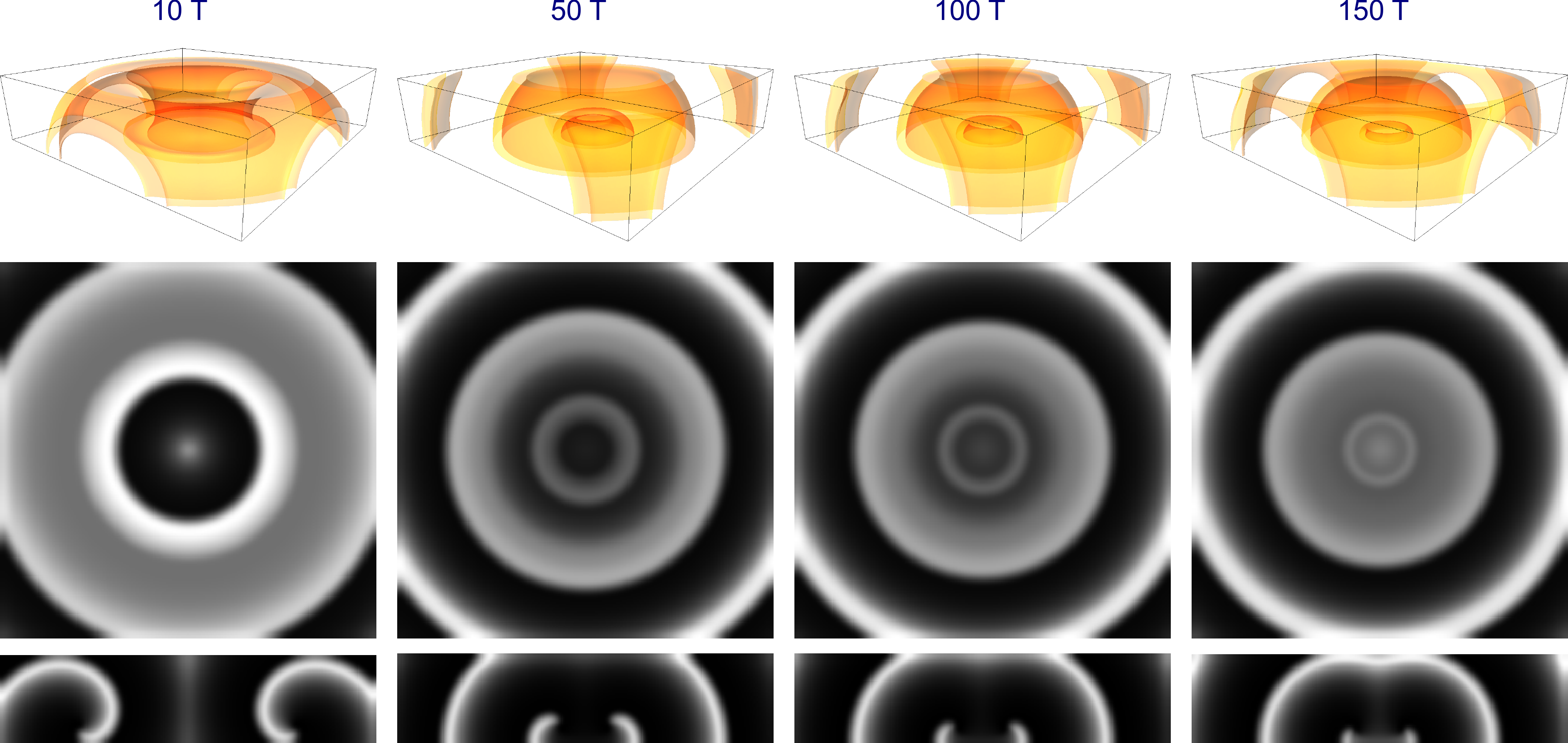}
\par\end{centering}

\caption{(Color online) Four snapshots of the wave pattern used for the results presented in Fig. \ref{fig:BoundZandRoverTime-Par2}. From top to bottom perspective, top, and side view as explained in the text. \label{fig:SRConfinedMultView}}
\end{figure*}

Snapshots of the emerging boundary-stabilized SR are presented in Fig. \ref{fig:SRConfinedMultView} where the 3d wavefronts (perspective view) are shown together with the corresponding radial cross sections (side view) and the pattern expected to be observed in the chemical experiment in a transparent gel layer (top view). The top view is obtained by summing up the local concentration values of the reduced catalyst (variable $v$ in the MCO model) along the $z$-direction. In the chemical experiment z-direction is the viewing direction, and the $v$ concentration determines the differences in the local brightness that are used to vizualize the pattern in transmitted light. Note that perspective and side view are not accessible in the chemical experiment 
without computer-tomographic imaging.

\section{Feasibility constraints for chemical experiments in a PBZ layer\label{sec:Height_Vs_Lightintensity}}

The reproducible optical initiation of SWs in thin PBZ layers opens promising possibilities for a thorough experimental analysis of their interaction with confining boundaries. 

In numerical experiments with the MCO model for the PBZ reaction we have studied SR initiation in realistic intervals of layer thickness, $d$, and intensity of incident light $\phi_{e}$ to estimate values of those parameters for which SR initiation seems feasible. In the corresponding 3d simulations, attenuation of the control light intensity with layer depth in $z$-direction according to Lambert-Beer's law 

\begin{equation}
	\phi(z)=\phi_{e}\exp\left(-\mu z\right),\label{eq:Lambert-Beer}
\end{equation}

was taken into account \cite{Amemiya:PRL77:1996,Amemiya:Chaos:98}. Parameter $\mu$ is the absorption coefficient and parameter $\phi_{e}$ the quantum efficiency for the photochemical bromide production \cite{Amemiya:PRL77:1996}. 

In the following we kept fixed $\mu=0.05$, and an illumination duration of approximately half of the spiral rotation period. Layer thickness $d$ and incident intensity $\phi_{e}$ were used as bifurcation parameters.

\begin{figure}
\begin{centering}
\includegraphics[width=1\columnwidth]{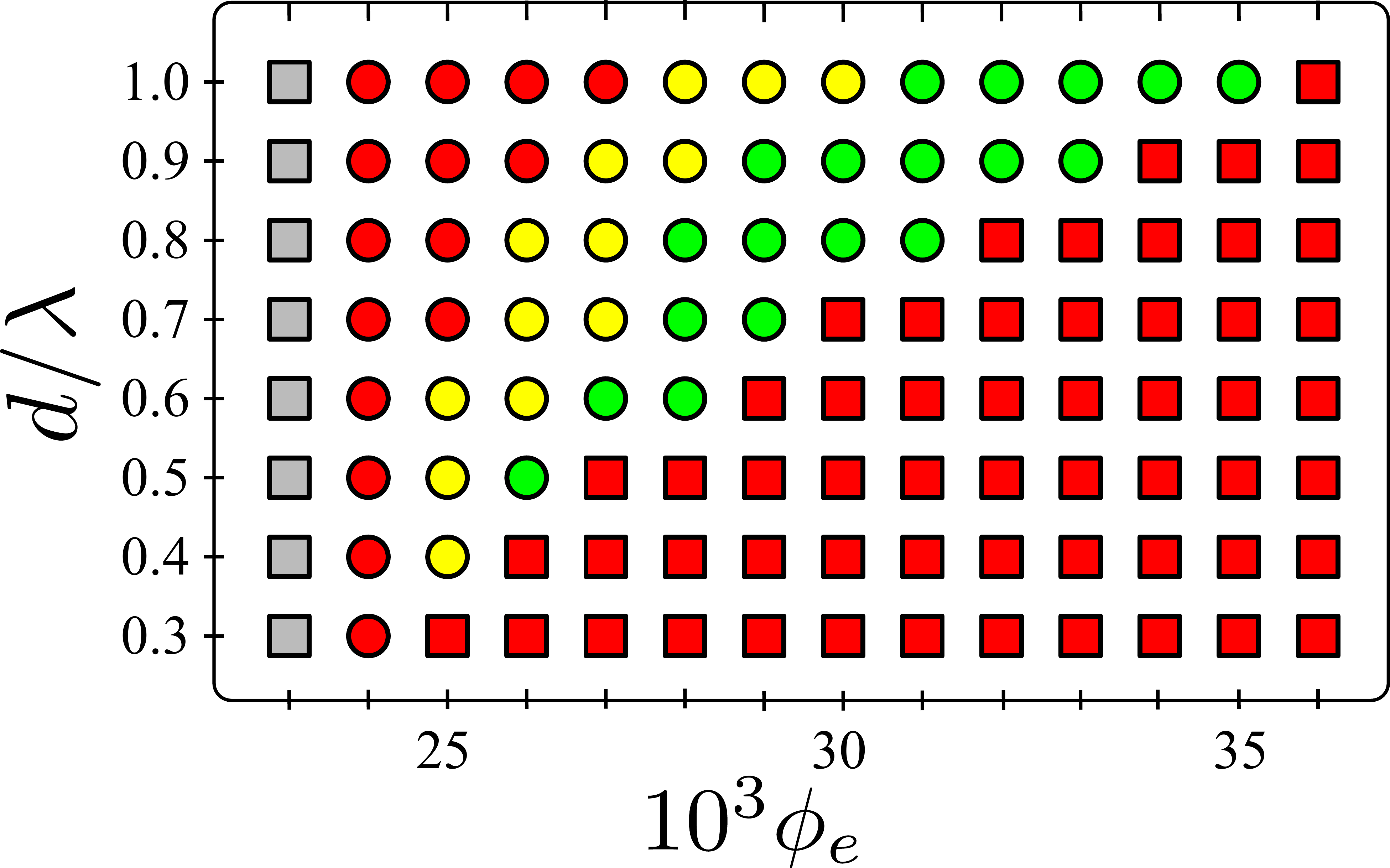}
\par\end{centering}

\caption{(Color online) Parameter plane spanned by the layer thickness in units of the spiral wave length and the intensity of incident light. Shown are numerical results for SR initiation by partial inhibition of cylindrical wave fronts in response to photochemically induced bromide production described by Lambert-Beer's law \eqref{eq:Lambert-Beer} with $\mu=0.05$.	
Grey and red squares: No SR initiation because of too weak and too strong photoinhibition of waves, respectively. Red and green circles: Successful initiation of SRs interacting with the NFB in  cooperative and antagonistic setting, respectively. Yellow circles: SR initiated far from the NFB. \label{fig:Phasediagram}}
\end{figure}

The results are summarized in the phase diagram shown in Fig. \ref{fig:Phasediagram}. Squares mark parameter regions where no SRs could be initiated: Below an incident light intensity of $\phi_{e}=24\cdot 10^{-3}$ the overall illumination is too weak to initiate stable SRs (grey squares), a layer thickness below $0.3\,\lambda$ is too narrow to support propagation of SRs, and red squares correspond to parameter combinations for which any wave acitivity in the layer is suppressed. In contrary, circles correspond to parameter regions with successful initiation of SRs. These SRs can interact cooperatively or antagonistically with the layer boundary (red and green circles, respectively). SRs initiated in the middle of the layer drift towards one of the layer boundaries (yellow circles). Their future evolution depends on whether they approach the boundary in the cooperative or the antagonistic setting. In summary, we found an extended parameter range with boundary-stabilized stationary SRs (green circles).

\begin{figure*}
\begin{centering}
\includegraphics[width=1\linewidth]{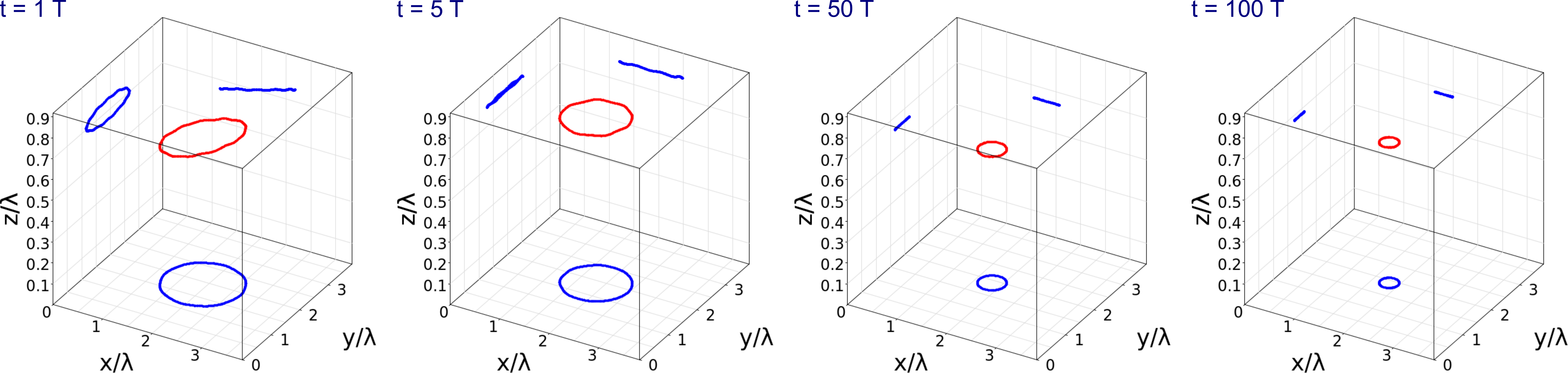}
\par\end{centering}

\caption{(Color online) Evolution of a SR filament (red) with projections in blue over 140 rotation periods $T$. Initially, the filament plane was inclined with respect to the confining NFB $z = 1$. Explanation see text.  Parameter values  $\phi_{e}=0.05$, $\mu=0.1$, $\phi_{i}=0.01$, and $L=125$ in \eqref{eq:inclining_function}. MCO parameters as in 
Fig. \ref{fig:SRFreeMultView}. \label{fig:FilamentInclined_Par02}}
\end{figure*}

So far we did not account for the attenuation of the background illumination treating parameter  $\phi_{0}$ as a constant, $\phi_{0}=0.014$. Actually, the background illumination decreases also in z-direction because of Lambert-Beer's law. Therefore, the SR evolves in the presence of a permanent excitability gradient in z-direction caused by the background illumination. This excitability gradient could influence the stability of the emerging SR. To check this possibility, we have numerically simulated SR evolution with z-dependent photochemically-induced bromide gradient given by  $\phi(z) = 0.014 \, e^{-\mu z} $. For three different values of the absorption coefficient, $\mu=0.0033$, $0.01$, and $0.25$, the SRs developed unaffected and proved to remain stable over more than $400$ rotation periods.

In the chemical experiment sometimes illumination during partial inhibition of the cylindrical wave is not spatially uniform in the $x$-$y$-plane. For example, applying an intensity gradient in $x$-direction 

\begin{equation}
\phi\left(x,z\right)=\phi_{0}+\phi_{e}\exp\left(-\mu z\right)+\frac{\phi_{i}}{L}x,\label{eq:inclining_function}
\end{equation}

the filament plane of the initiated SR wil not be parallel to the NFB $z = 0$. We have set $\phi_{e}=0.05$ and simulated the evolution of SR in antagonistic setting with filament plane inclined to the NFB for three different values $L = 57$, $100$ and $125$. After a short transient period of approximately five periods of spiral rotation the filament plane aligned in parallel to the NFB, and a stable SR developed. For $L=125$ this development is shown in Fig. \ref{fig:FilamentInclined_Par02}.

\section{Conclusion\label{sec:Conclusion}}

From our numerical findings we conclude that experiments performed in a thin transparent gel layer loaded with a photosensitive BZ catalyst are well suited in order to reveal 
the interaction of SWs with confining no-flux boundaries. Spatial confinement is an important issue in the dynamics of SWs as in many cases it can have a huge impact on 
the evolution in time and on the stability of the vortices \cite{Totz:2015, azhand:epl:2014, Dierckx:PRL:2012, Biktasheva:PRL:2015, Ke:Chaos:2015}.

Our calculations show, that for experimentally realistic conditions regarding the chemical composition, layer thickness and applied light intensities, axis-symmetric scroll rings can be 
initiated at a desired distance between filament plane and no-flux boundary. The preparation of initial conditions by self-completion of an optically cut cylindrical chemical wave greatly 
facilitates reproducible experiments under well-controlled laboratory conditions.

Moreover, our results point into another interesting direction. Obviously, for the correct interpretation of experimental results on wave propagation obtained in open gel reactors of the 
BZ reaction, three-dimensional effects might play an important role already for very thin gel layers. One example is non-annihilative head-on-collision of chemical waves \cite{Munuzuri:PRL:1997}.

\section{Acknowledgements}

This work was supported by the DFG in the framework of the collaborative research center 910.

\end{document}